\documentclass[pre,twocolumn,superscriptaddress,
showkeys,showpacs,amsmath,amsfonts,final]{revtex4}

\newif\ifpdf 
\ifx\pdfoutput\undefined
\pdffalse 
\else
\pdfoutput=1 
\pdftrue
\fi

\ifpdf 
\usepackage[pdftex]{graphicx}
\else 
\usepackage{graphicx}
\fi
\usepackage[cdot,squaren,textstyle]{SIunits}
\usepackage[sort&compress]{natbib}

\begin{document}

\title{Entropy and information in neural spike trains: Progress on the
  sampling problem}

\author{Ilya Nemenman}\email{nemenman@kitp.ucsb.edu}
\affiliation{Kavli Institute for Theoretical Physics, University of
  California, Santa Barbara, California 93106}

\author{William Bialek} \email{wbialek@princeton.edu}
\affiliation{Departments of Physics and the Lewis--Sigler Institute
  for Integrative Genomics, Princeton University, Princeton, New
  Jersey 08544}

\author{Rob de Ruyter van Steveninck} \email{deruyter@indiana.edu}
\affiliation{Department of Molecular Biology, Princeton University,
  Princeton, New Jersey 08544} 

\altaffiliation[Current address: ]{Department of Physics, Indiana
  University, 727 E. Third St., Bloomington, Indiana 47405}

\begin{abstract}
  The major problem in information theoretic analysis of neural
  responses and other biological data is the reliable estimation of
  entropy--like quantities from small samples. We apply a recently
  introduced Bayesian entropy estimator to synthetic data inspired by
  experiments, and to real experimental spike trains. The estimator
  performs admirably even very deep in the undersampled regime, where
  other techniques fail.  This opens new possibilities for the
  information theoretic analysis of experiments, and may be of general
  interest as an example of learning from limited data.
\end{abstract}



\preprint{NSF-KITP-03-43}
\maketitle

\section{Introduction}
\label{intro}

There has been considerable progress in using information theoretic
methods to sharpen and to answer many questions about the structure of
the neural code
\cite{BialekEtAl1991,TheunissenAndMiller1991,berry-97,strong-98,BorstAndTheunissen1999,brenner-00a,ReinagelAndReid2000,reich-01}.
Where classical experimental approaches have focused on mean response
of neurons to relatively simple stimuli, information theoretic methods
have the power to quantify the responses to arbitrarily complex and
even fully natural stimuli \cite{spikes,LewenEtAl2001}, taking account
of both the mean response and its variability in a rigorous way,
independent of detailed modeling assumptions.  Measurements of entropy
and information in spike trains also allow us to test directly the
hypothesis that the neural code adapts to the distribution of sensory
inputs, optimizing the rate or efficiency of information transmission
\cite{barlow-59,barlow-61,laughlin-81,BrennerEtAl2000,FairhallEtAl2001}.

A problem with such measurements is that entropy and information
depend explicitly on the full distribution of neural responses, just a
limited sample of which is provided by experiments. In particular, we
need to know the distribution of responses to each stimulus in our
ensemble, and the number of samples from this distribution is limited
by the number of times the full set of stimuli can be repeated.  For
natural stimuli with long correlation times the time required to
present a useful ``full set of stimuli'' is long, limiting the number
of independent samples we can obtain from stable neural recordings.
Furthermore, natural stimuli generate neural responses of high timing
precision, and thus the space of meaningful responses itself is very
large \cite{mainen-95,rds-97,berry-97,LewenEtAl2001}.  These factors
make the sampling problem more serious as we move to more interesting
and natural stimuli.

A natural response to this problem is to give up the generality of a
completely model independent information theoretic approach.  Some
explicit help from models is required to regularize learning of the
underlying probability distributions from the experiments.  The
question is if we can keep the generality of our analysis by
introducing the gentlest of regularizations for the abstract learning
problem, or if we need stronger assumptions about the structure of the
neural code itself (for example, introducing a metric on the space of
responses \cite{victor-purpura-97,Victor2002}).

A classical problem suggests that we may succeed even with very weak
assumptions. Remember that one needs to have only $N\sim 23$ people in
a room before any two of them are reasonably likely to share the same
birthday. This is much less than $K=365$, the number of possible
birthdays.  Turning this around, we can estimate the number of
possible birthdays by polling $N$ people and counting how often we
find coincidences.  Once $N$ is large enough to have observed a few of
those, we can get a pretty good estimate of $K$.  This will happen
with a significant probability for $N \sim \sqrt{K}\ll K$.

The idea of estimating entropy by counting coincidences was proposed
long ago by Ma \cite{ma-81} for physical systems in the microcanonical
ensemble where distributions should be uniform at fixed energy.
Clearly, if we could generalize the Ma idea to arbitrary
distributions, then we would be able to explore a much wider variety
of question about information in the neural code.  Here we argue that
a simple and abstract Bayesian prior, introduced in Ref.~\cite{nsb},
comes close to the objective.

It is well known that, for $N<K$, there are no universally good
entropy estimators \cite{paninski-03,rubinfeld-02}.  Thus the main
question is: does a particular method work well only for (possibly
irrelevant) abstract model problems, or can it also be trusted for
natural data?  Hence our goal is neither to search for potential
theoretical limitations of the approach (these must exist and have
been found), nor to analyze the neural code (this will be left for the
future).  Instead we aim at convincingly showing that the method of
Ref.~\cite{nsb} can generate reliable estimates of entropy well into a
classically undersampled regime for an experimentally relevant case of
neurophysiological recordings.

\section{An estimation strategy}
\label{strategy}
Consider the problem of estimating the entropy $S$ of a probability
distribution $\{p_{\rm i}\}$, $ S = -\sum_{{\rm i}=1}^K p_{\rm
  i}\log_2 p_{\rm i}$,  where the index $\rm i$ runs over $K$
possibilities (e.g., $K$ possible neural responses). In an experiment
we observe that in $N$ examples each possibility $\rm i$ occurred
$n_{\rm i}$ times.  If $N \gg K$, we approximate the probabilities by
frequencies, $p_{\rm i} \approx f_{\rm i} \equiv n_{\rm i} /N$, and
construct a naive estimate of the entropy,
\begin{equation}
S_{\rm naive} = -\sum_{{\rm i}=1}^K f_{\rm i}\log_2 f_{\rm i} .
\end{equation}
This is also a maximum likelihood estimator, since the maximum
likelihood estimate of the probabilities is given by the frequencies.
Thus we will replace $S_{\rm naive}$ by $S^{\rm ML}$ in what follows.

It is well know that $S^{\rm ML}$ underestimates the entropy (cf.\ 
Ref.~\cite{paninski-03}).  With good sampling ($N \gg K$), classical
arguments due to Miller \cite{miller-55} show that the ML estimate
should be corrected by a universal term $(K-1)/2N$, and several groups
have used this correction in the analysis of neural data.  In
practice, many bins may have truly zero probability (for example, as a
result of refractoriness; see below), and the samples from the
distribution might not be completely independent.  Then $S^{\rm ML}$
still deviates from the correct answer by a term $\propto 1/N$, but
the coefficient is no longer known a priori. Under these conditions
one can heuristically verify and extrapolate the $1/N$ behavior from
subsets of the available data \cite{strong-98}.  Alternatively, still
agreeing on the $1/N$ correction, one can calculate its coefficient
(interpretable as an effective number of bins $K^*$) for some classes
of distributions
\cite{grassberger-88,panzeri-treves-96,grassberger-03}.  All of these
approaches, however, work only when the sampling errors are in some
sense a small perturbation.

If we want to make progress outside of the asymptotically large $N$
regime we need an estimator that does not have a perturbative
expansion in $1/N$ with $S_{\rm ML}$ as the zeroth order term. The
estimator of Ref.~\cite{nsb} has just this property. Recall that
$S_{\rm ML}$ is a limiting case of Bayesian estimation with Dirichlet
priors.  Formally, we consider that the probability distributions
${\bf p} \equiv \{p_{\rm i}\}$ are themselves drawn from a
distribution ${\mathcal P}_\beta ({\bf p})$ of the form
\begin{equation}
{\mathcal P}_\beta ({\bf p}) = \frac{1}{Z(\beta; K)}
\left[\prod_{{\rm i}=1}^K p_{\rm i}^{(\beta-1)}\right]
 \delta \Bigl( \sum_{{\rm i}=1}^K p_{\rm i} -1 \Bigr) ,
\end{equation}
where the delta function enforces normalization of distributions ${\bf
  p}$ and the partition function $Z(\beta ; K)$ normalizes the prior
${\mathcal P}_\beta ({\bf p})$.  Maximum likelihood estimation is
Bayesian estimation with this prior in the limit $\beta \rightarrow
0$, while the natural ``uniform'' prior is $\beta =1$.  The key
observation of Ref.~\cite{nsb} is that while these priors are quite
smooth on the space of ${\bf p}$, the distributions drawn at random
from ${\mathcal P}_\beta$ all have very similar entropies, with a
variance that vanishes as $K$ becomes large.  Fundamentally, this is
the origin of the sample size dependent bias in entropy estimation,
and one might thus hope to correct the bias at its source.  The goal
then is to construct a prior on the space of probability distributions
which generates a nearly uniform distribution of entropies.  Because
the entropy of distributions chosen from ${\mathcal P}_\beta$ is
sharply defined {\em and} monotonically dependent on the parameter
$\beta$, we can come close to this goal by an average over $\beta$,
\begin{equation}
  {\mathcal P}_{\rm NSB}  ({\bf p} )  \propto \int d\beta 
   \,\frac{d \bar S(\beta;K)}{d\beta}\,
  {\mathcal P}_\beta ({\bf p})\,.
\end{equation}
Here $\bar S(\beta;K)$ is the average entropy of distributions chosen
from ${\mathcal P}_\beta$ \cite{ww,nsb},
\begin{equation}
  \bar S(\beta;K)  \equiv \xi =
  \psi_0(K\beta+1) 
  -\psi_0(\beta+1) \, ,
  \label{Sap}
\end{equation}
where $\psi_m(x) = (d/dx)^{m+1} \log_2 \Gamma(x)$ are the polygamma
functions.

Given this prior, we proceed in standard Bayesian fashion.  The
probability of observing the data ${\bf n}\equiv\{n_{\rm i}\}$ given
the distribution $\bf p$ is
\begin{equation}
P({\bf n} | {\bf p}) \propto \prod_{{\rm i}=1}^K p_{\rm i}^{n_{\rm i}} ,
\end{equation}
and then
\begin{eqnarray}
  P({\bf p} | {\bf n}) &=& 
  P({\bf n} | {\bf p})   {\mathcal P}_{\rm NSB}  ({\bf p} ){\bf \cdot}
  \frac{1}{P({\bf n})},\\
  P({\bf n}) &=& \int  d{\bf p} \,P({\bf n} | {\bf p})
  {\mathcal P}_{\rm NSB}  ({\bf p}
  ),\\
  \left(S^{\rm NSB}\right)^m &=& \int  d{\bf p} \,
  \left( -\sum_{{\rm i=1}}^K
    p_{\rm i} \log_2 p_{\rm i} \right)^m P({\bf p} | {\bf n}) .
\end{eqnarray}
Here we need to calculate the first two posterior moments of the
entropy, $m=1,2$, in order to have an access to the entropy estimate
and to its variance as well.

The Dirichlet priors allow all the ($K$ dimensional) integrals over
$\bf p$ to be done analytically, so that the computation of $S^{\rm
  NSB}$ and of its posterior error reduces to just three numerical
one--dimensional integrals:
\begin{eqnarray}
  \left(S^{\rm NSB}\right)^m &=& \frac{\int d\xi\, 
    \rho(\xi,{\bf n}) \, S_\beta^m ({\bf n})}
  {\int d\xi\, \rho(\xi,{\bf n})}\,,\;\;\;\mbox{where}
  \label{Shat}
  \\
  \rho(\xi, {\bf n}) &=&
  \frac{\Gamma(K\beta(\xi))}{\Gamma(N+K\beta(\xi))}\,
  \prod_{i=1}^K \frac{\Gamma(n_i+\beta(\xi))}{\Gamma(\beta(\xi))}\,,
  \label{rho}
\end{eqnarray}
where the one--to--one relation between $\beta$ and $\xi$ is given by
Eq.~(\ref{Sap}), and $S_\beta^m({\bf n})$ is the expectation value of
the $m$-th entropy moment at fixed $\beta$; the exact expression for
$m=1,2$ is given in Ref.~\cite{ww}.

Details of the NSB method can be found in Refs.~\cite{nsb,nsb2}, and
the source code of the implementations in either Octave/C++ or plain
C++ is available from the authors.  We draw attention to several
points.

First, since the analysis is Bayesian, we obtain not only $S^{\rm NSB}$
but also its a posteriori standard deviation, $\delta S^{\rm
  NSB}$---an error bar on our estimate, see Eq.~(\ref{Shat}).

Second, for $N\to\infty$ and $N/K\to 0$ the estimator admits
asymptotic analysis. The important parameter is the number of
coincidences $\Delta = N-K_1$, where $K_1$ is the number of bins with
non-zero counts. If $\Delta/N\to {\rm const}<1$ (many coincidences),
then the standard saddle point evaluation of the integrals in
Eq.~(\ref{Sap}) is possible. Interestingly, the second derivative at
the saddle is $(\ln^2 2)\, \Delta$ to the leading order in $\Delta/N$.
The second asymptotic can be obtained for $\Delta\sim O(N^0)$ (few
coincidences).  Then
\begin{eqnarray}
  S^{\rm NSB} &\approx&\frac{C_\gamma}{\ln 2} - 1 + 2 \log_2 N
  -\psi_0(\Delta)\,,
  \label{Shat_res}
  \\ 
  \delta S^{\rm NSB} &\approx& \sqrt{\psi_1(\Delta)}\,,
  \label{dShat_res}
\end{eqnarray}
where $C_\gamma$ is the Euler's constant. This is particularly
interesting since $S^{\rm NSB}$ happens to have a finite limit for
$K\to\infty$, thus allowing entropy estimation even for infinite (or
unknown) cardinalities.

Third, both of the above asymptotics show that the estimation
procedure relies on $\Delta$ to make its estimates; this is in the
spirit of Ref.~\cite{ma-81}.

Finally, $S^{\rm NSB}$ is unbiased if the distribution being learned
is typical in ${\mathcal P}_\beta({\bf p})$ for some $\beta$, that is,
its rank ordered (Zipf) plot is of the form
\begin{eqnarray}
q_i &\approx& 1 - \left[\frac{ \beta B(\beta, K\beta - \beta )  (K-1) \,i}
{K} \right] ^{1/(K\beta-\beta)}, 
\label{left}\\
q_i &\approx& \left[ \frac{ \beta B(\beta, K\beta - \beta )  (K-i+1)}
{K}\right]^{1/\beta},
\label{right}
\end{eqnarray}  
for $i/K\to 0$ and $i/K\to1$ respectively. If the Zipf plot has tails
that are too short (too long), then the estimator should over (under)
estimate.  While underestimation may be severe (though always strictly
smaller than that for $S^{\rm ML}$), overestimation is very mild, if
present at all, in the most interesting regime $1\ll\Delta\ll N$.
$S^{\rm NSB}$ is also unbiased for distributions that are typical in
some weighted combinations of ${\mathcal P}_\beta$ for different
$\beta$'s, in particular in ${\mathcal P}_{\rm NSB}$ itself. However,
the typical Zipf plots in this case are more complicated and will be
detailed elsewhere.

Before proceeding it is worth asking what we hope to accomplish. Any
reasonable estimator will converge to the right answer in the limit of
large $N$.  In particular, this is true for $S^{\rm NSB}$, which is a
{\em consistent} Bayesian estimator \footnote{In reference to Bayesian
  estimators, consistency usually means that, as $N$ grows, the
  posterior probability concentrates around unknown parameters of the
  true model that generated the data. For finite parameter models,
  such as the one considered here, only technical assumptions like
  positivity of the prior for all parameter values, soundness
  (different parameters always correspond to different distributions)
  \cite{clarke-barron-90}, and a few others are needed for
  consistency.  For nonparametric models, the situation is more
  complicated. There one also needs ultraviolet convergence of the
  functional integrals defined by the prior
  \cite{nemenman-00,bnt-01}.}.  The central problem of entropy
estimation is systematic bias, which will cause us to (perhaps
significantly) under- or overestimate the information content of spike
trains or the efficiency of the neural code. The bias, which vanishes
for $N\to\infty$, will manifest itself as a systematic drift in plots
of the estimated value versus the sample size. A successful estimator
would remove this bias as much as possible.  Ideally we thus hope to
see an estimate which for all values of $N$ is within its error bars
from the correct answer.  As $N$ increases the error bars should
narrow, with relatively little variation of the (mean) estimate
itself. When data are such that no reliable estimation is possible,
the estimator should remain uncertain, that is, the posterior variance
should be large. The main purpose of this paper is to show that the
NSB procedure applied to natural and nature--inspired synthetic
signals comes close to this ideal over a wide range of $N \ll K$, and
even $N \ll 2^S$.  The procedure thus is a viable tool for
experimental analysis.

\section{A model problem}
\label{modprob}

It is important to test our techniques on a problem which captures
some aspects of real world data yet is sufficiently well defined that
we know the correct answer.  We constructed synthetic spike trains
where intervals between successive spikes were independent and chosen
from an exponential distribution with a dead time or refractory period
of $g=1.8$ \milli\second; the mean spike rate was $r=0.26$
spikes/\milli\second.  This corresponds to the rate of $r_0 = r/(1 -
rg) = 0.49$ spikes/\milli\second\ for the part of the signal, where
spiking is not prohibited by refractoriness.  These parameters are
typical of the high spike rate, noisy regions of the experiment
discussed below, which provide the greatest challenge for entropy
estimation.

Following the scheme outlined in Ref.~\cite{strong-98}, we examine the
spike train in windows of duration $T=15$ \milli\second\ and
discretize the response with a time resolution $\tau = 0.5$
\milli\second.  Because of the refractory period each bin of size
$\tau$ can contain at most one spike, and hence the neural response is
a binary word with $T/\tau = 30$ letters.  The space of responses has
$K = 2^{30}\approx 10^9$ possibilities.  Of course, most of these have
probability exactly zero because of refractoriness, and the number of
possible responses consistent with this constraint is bounded by $\sim
2^{16} \approx 10^5$.  An approximation to the entropy of this
distribution, is given by an appropriate correction to Eq.~(3.21) of
Ref.~\cite{spikes}, the entropy of a non--refractory Poisson process:
\begin{equation}
S =\frac{rT}{\ln 2} \left[-\ln \left(1-{\rm e}^{-r_0\tau} \right) 
  + \frac{r_0\tau\, {\rm e}^{-r_0\tau}}{1-{\rm e}^{-r_0\tau}}\right]=
 13.57~{\rm bits}.
\end{equation}

\begin{figure}
  \ifpdf \includegraphics[width=3.2in]{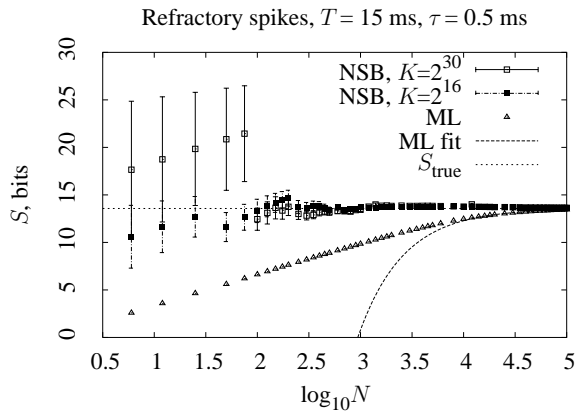} \else
  \includegraphics[height=3.2in, angle=270]{refractory} \fi
  \caption{\label{fig:artif}Entropy estimation for a model
    problem. Notice that the estimator reaches the true value within
    the error bars as soon as $N^2 \sim 2^S$, at which point
    coincidences start to occur with high probability. Slight
    overestimation for $N>10^3$ is expected (see text) since this
    distribution is atypical in ${\mathcal P}_{\rm NSB}$.}
\end{figure}
In Fig.~\ref{fig:artif} we show the results of entropy estimation for
this model problem. As expected, the naive estimate $S^{\rm ML}$
reaches its asymptotic behavior only when $N > 2^S$, thus the $1/N$
extrapolation becomes successful at $N\sim10^4$ (the ``ML fit'' line
on the plot).  In contrast, we see that $S^{\rm NSB}$ gives the right
answer within errors at $N \sim 100$.  We can improve convergence by
providing the estimator with the ``hint'' that the number of possible
responses $K$ is much smaller than the upper limit of $2^{30}$, but
even without this hint we have excellent entropy estimates already at
$N \sim (2^S)^{1/2}$.  This is in accord with expectations from Ma's
analysis of (microcanonical) entropy estimation \cite{ma-81}. However,
here we achieve these results for a nonuniform distribution.

\section{Analyzing real data}

\begin{figure}
  \ifpdf \includegraphics[width=3in]{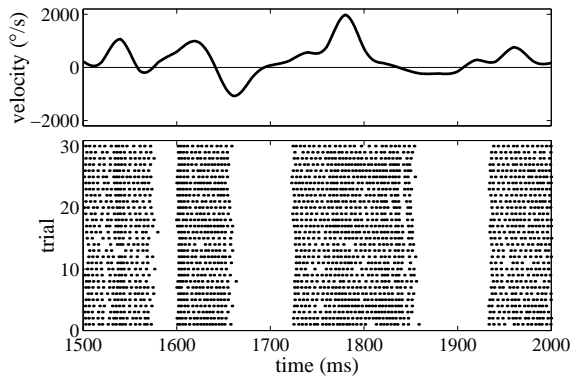}
  \else \includegraphics[width=3in,angle=0]{sfly022_vel_raster} \fi
  \caption{\label{fig:flyexp} Data from a fly motion sensitive neuron
    in a natural stimulus setting. Top: a 500 \milli\second\ section
    of a 10 \second\ angular velocity trace that was repeated 196
    times.  Bottom: raster plot showing the
    response to 30 consecutive trials; each dot marks the occurrence of a spike.}
\end{figure}

For a test on real neurophysiological data, we use recordings from a
wide field motion sensitive neuron (H1) in the visual system of the
blowfly {\em Calliphora vicina}. While action potentials from H1 were
recorded, the fly rotated on a stepper motor outside among the bushes,
with time dependent angular velocity representative of natural flight.
Figure~\ref{fig:flyexp} presents a sample of raw data from such an
experiment (see~Ref.~\cite{LewenEtAl2001} for details).

Following Ref.~\cite{strong-98}, the information content of a spike
train is the difference between its total entropy and the entropy of
neural responses to repeated presentations of the same stimulus
\footnote{It may happen that information is a small difference between
  two large entropies. Then, due to statistical errors, methods that
  estimate information directly will have an advantage over NSB, which
  estimates entropies first. In our case, this is not a problem since
  the information is roughly a half of the total available entropy
  \cite{strong-98}.}. The latter is substantially more difficult to
estimate. It is called the noise entropy $S_n$, since it measures
response variations that are uncorrelated with the sensory input. The
noise in neurons depends on the stimulus itself---there are, for
example, stimuli which generate with certainty zero spikes in a given
window of time---and so we write $S_{n|t}$ to mark the dependence on
the time $t$ at which we take a slice through the raster of responses.
In this experiment the full stimulus was repeated 196 times, which
actually is a relatively large number by the standards of
neurophysiology.  The fly makes behavioral decisions based on $\sim
10- 30~\milli\second$ windows of its visual input
\cite{LandAndCollett1974}, and under natural conditions the time
resolution of the neural responses is of order 1 \milli\second\ or
even less \cite{LewenEtAl2001}, so that a meaningful analysis of
neural responses must deal with binary words of length $10-30$ or
more. Refractoriness limits the number of these words which can occur
with nonzero probability (as in our model problem), but nonetheless we
easily reach the limit where the number of samples is substantially
smaller than the number of possible responses.

\begin{figure}[t]
   \ifpdf
   \includegraphics[width=2.7in]{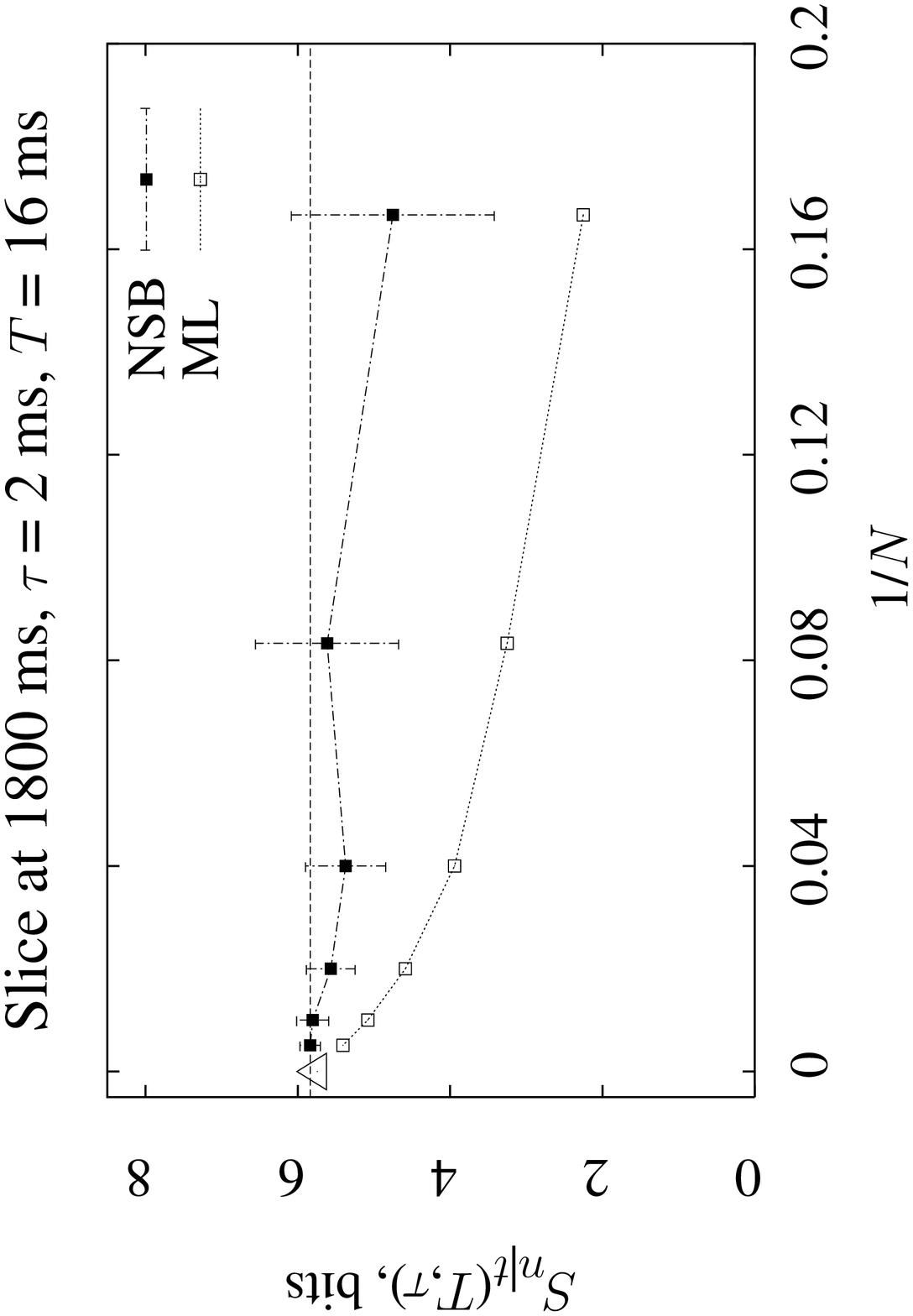}
   \includegraphics[width=2.7in]{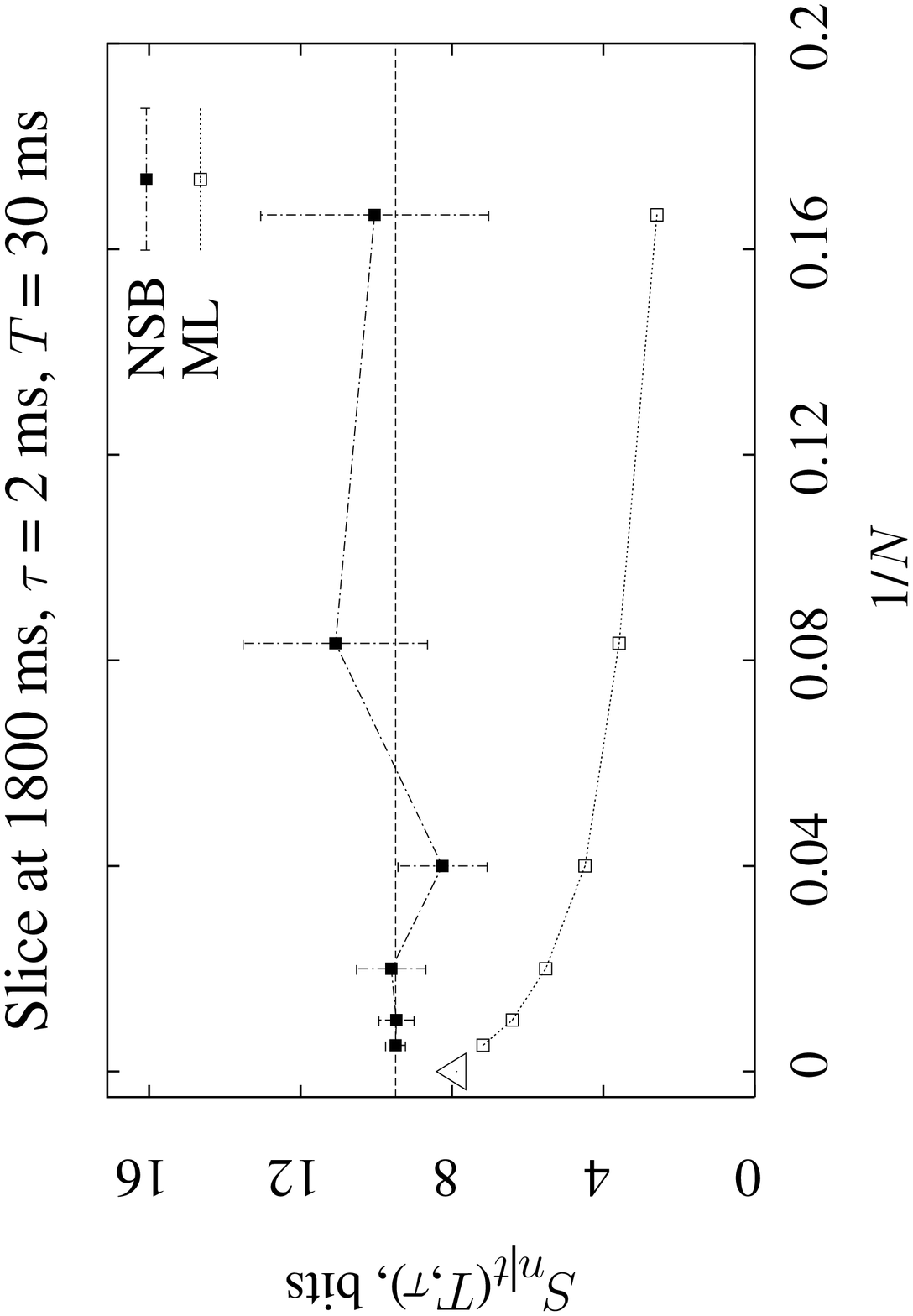}
   \else
   \includegraphics[height=2.7in,angle=270]{T8_nsb_vs_ml_2}
   \includegraphics[height=2.7in,angle=270]{T15_nsb_vs_ml_2}
   \fi
   \caption{\label{fig:nsbml}Slice entropy vs.\ sample size.  Dashed
     line on both plots is drawn at the value of $\left.S^{\rm
         NSB}\right|_{N=N_{\rm max}}$ to show that the estimator is
     stable within its error bars even for very low $N$. Triangle
     corresponds to the value of $S^{\rm ML}$ extrapolated to
     $N\to\infty$ from the four largest values of $N$. First and second
     panels show examples of word lengths for which $S_{\rm ML}$ can
     or cannot be reliably extrapolated. $S^{\rm NSB}$ is stable in
     both cases, shows no $N$ dependent drift, and agrees with $S^{\rm
       ML}$ where the latter is reliable.}
\end{figure}
Let us start by looking at a single moment in time,
$t=1800~\milli\second$ from the start of the repeated stimulus, as in
Fig.~\ref{fig:flyexp}.  If we consider a window of duration $T =
16~\milli\second$ at time resolution $\tau = 2~\milli\second$
\footnote{For our and many other neural systems, the spike timing can
  be more accurate than the refractory period of roughly 2
  \milli\second\ \cite{brenner-00a,rob-01,LewenEtAl2001}.  For the
  current amount of data, discretization of $\tau\ll1~\milli\second$
  and large enough $T$ will push the limits of all estimation methods,
  including ours, that do not make explicit assumptions about
  properties of the spike trains. Thus, to have enough statistics to
  convincingly show validity of the NSB approach, in this paper we
  choose $\tau =0.75\dots2~\milli\second$, which is still much shorter
  than other methods can handle. We leave open the possibility that
  more information is contained in timing precision at finer scales.},
we obtain the entropy estimates shown in the first panel of
Fig.~\ref{fig:nsbml}.  Notice that in this case we actually have a
total number of samples which is comparable to or larger than
$2^{S_{n|t}}$, and so the maximum likelihood estimate of the entropy
is converging with the expected $1/N$ behavior.  The NSB estimate
agrees with this extrapolation.  The crucial result is that the NSB
estimate is correct within error bars across the whole range of $N$;
there is a slight variation in the mean estimate, but the main effect
as we add samples is that the error bars narrow around the correct
answer.  In this case our estimation procedure has removed essentially
all of the sample size dependent bias.

As we open our window to $T = 30~\milli\second$, the number of
possible responses (even considering refractoriness) is vastly larger
than the number of samples. As we see in the second panel of
Fig.~\ref{fig:nsbml}, any attempt to extrapolate the ML estimate of
entropy now requires some wishful thinking.  Nonetheless, in parallel
with our results for the model problem, we find that the NSB estimate
is stable within error bars across the full range of available $N$.

\begin{figure}
  \centerline{\includegraphics[width=2.9in,height=2.3in]{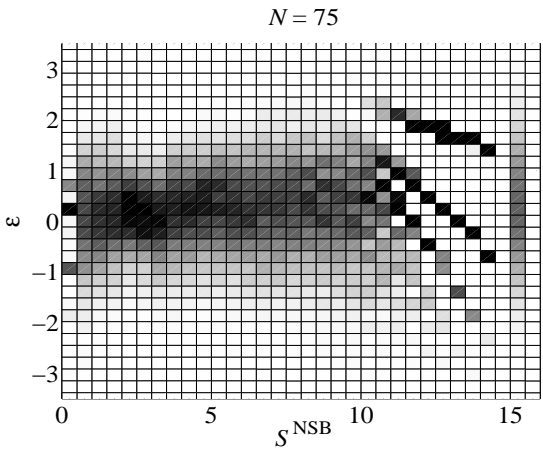}}
  \caption{\label{fig:conds}Distribution of the normalized entropy
    error conditional on $S^{\rm NSB}(N_{\rm max})$ for $N=75$ and
    $\tau=0.75~\milli\second$. Darker patches correspond to higher
    probability.  The band in the right part of the plot is the normal
    distribution around zero with the standard deviation of 1 (the
    standard deviation of plotted conditional distributions averaged
    over $S^{\rm NSB}$ is about 0.7, which indicates a non--Gaussian
    form of the posterior for small number of coincidences
    \cite{nsb2}). For values of $S^{\rm NSB}$ up to about 12 bits the
    estimator performs remarkably well. For yet larger entropies,
    where the number of coincidence is just a few, the discrete nature
    of the estimated values is evident, and this puts a bound on
    reliability of $S^{\rm NSB}$.}
\end{figure}

For small $T$ we can compare the results of our Bayesian estimation
with an extrapolation of the ML estimate; each moment in time relative
to the repeated stimulus provides an example.  We have found that the
results in the first panel of Fig. \ref{fig:nsbml} are typical: in the
regime where extrapolation of the ML estimator is reliable, our
estimator agrees within error bars over a broad range of sample sizes.
More precisely, if we take the extrapolated ML estimate as the correct
answer, and measure the deviation of $S^{\rm NSB}$ from this answer in
units of the predicted error bar, we find that the mean square value
of this normalized error is of order one.  This is as expected if our
estimation errors are random rather than systematic.

For larger $T$ we do not have a calibration against the (extrapolated)
$S^{\rm ML}$, but we can still ask if the estimator is stable, within
error bars, over a wide range of $N$.  To check this stability we
treat the value of $S^{\rm NSB}$ at $N=N_{\rm max}=196$ as our best
guess for the entropy and compute the normalized deviation of the
estimates at smaller values of $N$ from this guess,
$\varepsilon=\left[S^{\rm NSB}(N) - S^{\rm NSB}(N_{\rm
    max})\right]/\delta S^{\rm NSB}(N)$.  Again, each moment in time
is an example.  Figure~\ref{fig:conds} shows the distribution of these
normalized deviations conditional on the entropy estimate with $N=75$;
this analysis is done for $\tau = 0.75~\milli\second$, with $T$ in the
range between $1.5~\milli\second$ and $22.5~\milli\second$.  Since the
different time slices span a range of entropies, over some range we
have $N > 2^S$, and in this regime the entropy estimate must be
accurate (as in the analysis of small $T$ above).  Throughout this
range, the normalized deviations fall in a narrow band with mean close
to zero and a variance of order one, as expected if the only
variations with the sample size were random.  Remarkably this pattern
continues for larger entropies, $S > \log_2 N=6.2$ bits, demonstrating
that our estimator is stable even deep into the undersampled regime.
This is consistent with the results obtained in our model problem, but
here we find the same answer for the real data.

Note that Fig. \ref{fig:conds} illustrates results with $N$ less than
one half the total number of samples, so we really are testing for
stability over a large range in $N$.  This emphasizes that our
estimation procedure moves smoothly from the well sampled into the
undersampled regime without accumulating any clear signs of systematic
error.  The procedure collapses only when the entropy is so large that
the probability of observing the same response more than once (a
coincidence) becomes negligible.

\section{Discussion}


The estimator we have explored here is constructed from a prior that
has a nearly uniform distribution of entropies.  It is plausible that
such a uniform prior would largely remove the sample size dependent
bias in entropy estimation, but it is crucial to test this
experimentally. In particular, there are infinitely many priors which
are approximately (and even exactly) uniform in entropy, and it is not
clear which of them will allow successful estimation in real world
problems.  We have found that the NSB prior almost completely removed
the bias in the model problem (Fig.~\ref{fig:artif}). Further, for
real data in a regime where undersampling can be beaten down by data
the bias is removed to yield agreement with the extrapolated ML
estimator even at very small sample sizes (Fig.~\ref{fig:nsbml}, first
panel).  Finally and most crucially, the NSB estimation procedure
continues to perform smoothly and stably past the nominal sampling
limit of $N \sim 2^S$, all the way to the Ma cutoff $N^2 \sim 2^S$
(Fig.~\ref{fig:conds}).  This opens the opportunity for rigorous
analysis of entropy and information in spike trains under a much wider
set of experimental conditions.

\acknowledgments

We thank J Miller for important discussions, GD Lewen for his help
with the experiments, which were supported by the NEC Research
Institute, and the organizers of the NIC'03 workshop for providing a
venue for a preliminary presentation of this work.  IN was supported
by NSF Grant No.\ PHY99-07949 to the Kavli Institute for Theoretical
Physics.  IN is also very thankful to the developers of the following
Open Source software: GNU Emacs, GNU Octave, GNUplot, and te\TeX.

\bibliographystyle{unsrtnat} {\small\bibliography{flies}}

\end{document}